\documentstyle[epsfig]{mn}

\newif\ifAMStwofonts
\def\gtsima{$\; \buildrel > \over \sim \;$}
\def\ltsima{$\; \buildrel < \over \sim \;$}
\def\gsim{\lower.5ex\hbox{\gtsima}}
\def\lsim{\lower.5ex\hbox{\ltsima}}
\newcommand{\etal}{et al.\ }
\def\Msun{{M_\odot}}
\def\be{\begin{equation}}
\def\ee{\end{equation}}
\ifoldfss
  \ifCUPmtlplainloaded \else
    \NewTextAlphabet{textbfit} {cmbxti10} {}
    \NewTextAlphabet{textbfss} {cmssbx10} {}
    \NewMathAlphabet{mathbfit} {cmbxti10} {} % for math mode
    \NewMathAlphabet{mathbfss} {cmssbx10} {} %  "   "    "
  \fi
  \ifAMStwofonts
    \ifCUPmtlplainloaded \else
      \NewSymbolFont{upmath} {eurm10}
      \NewSymbolFont{AMSa} {msam10}
      \NewMathSymbol{\upi}     {0}{upmath}{19}
      \NewMathSymbol{\umu}     {0}{upmath}{16}
      \NewMathSymbol{\upartial}{0}{upmath}{40}
      \NewMathSymbol{\leqslant}{3}{AMSa}{36}
      \NewMathSymbol{\geqslant}{3}{AMSa}{3E}

      \let\leq=\leqslant \let\le=\leqslant
      \let\geq=\geqslant \let\ge=\geqslant
    \fi
  \fi
\fi % End of OFSS

\ifnfssone
  \newmathalphabet{\mathit}
  \addtoversion{normal}{\mathit}{cmr}{m}{it}
  \addtoversion{bold}{\mathit}{cmr}{bx}{it}
  \newmathalphabet{\mathbfit} % math mode version of \textbfit{..}
  \addtoversion{normal}{\mathbfit}{cmr}{bx}{it}
  \addtoversion{bold}{\mathbfit}{cmr}{bx}{it}
  \newmathalphabet{\mathbfss} % math mode version of \textbfss{..}
  \addtoversion{normal}{\mathbfss}{cmss}{bx}{n}
  \addtoversion{bold}{\mathbfss}{cmss}{bx}{n}
  \ifAMStwofonts
    \ifCUPmtlplainloaded \else
      \UseAMStwoboldmath
      \makeatletter
      \new@mathgroup\upmath@group
      \define@mathgroup\mv@normal\upmath@group{eur}{m}{n}
      \define@mathgroup\mv@bold\upmath@group{eur}{b}{n}
      \edef\UPM{\hexnumber\upmath@group}
      \new@mathgroup\amsa@group
      \define@mathgroup\mv@normal\amsa@group{msa}{m}{n}
      \define@mathgroup\mv@bold\amsa@group{msa}{m}{n}
      \edef\AMSa{\hexnumber\amsa@group}
      \makeatother
      \mathchardef\upi="0\UPM19
      \mathchardef\umu="0\UPM16
      \mathchardef\upartial="0\UPM40
      \mathchardef\leqslant="3\AMSa36
      \mathchardef\geqslant="3\AMSa3E

      \let\leq=\leqslant \let\le=\leqslant
      \let\geq=\geqslant \let\ge=\geqslant
    \fi
  \fi
\fi % End of NFSS release 1

\ifnfsstwo
  \DeclareMathAlphabet{\mathbfit}{OT1}{cmr}{bx}{it}
  \SetMathAlphabet\mathbfit{bold}{OT1}{cmr}{bx}{it}
  \DeclareMathAlphabet{\mathbfss}{OT1}{cmss}{bx}{n}
  \SetMathAlphabet\mathbfss{bold}{OT1}{cmss}{bx}{n}
  \ifAMStwofonts
    \ifCUPmtlplainloaded \else
      \DeclareSymbolFont{UPM}{U}{eur}{m}{n}
      \SetSymbolFont{UPM}{bold}{U}{eur}{b}{n}
      \DeclareSymbolFont{AMSa}{U}{msa}{m}{n}
      \DeclareMathSymbol{\upi}{0}{UPM}{"19}
      \DeclareMathSymbol{\umu}{0}{UPM}{"16}
      \DeclareMathSymbol{\upartial}{0}{UPM}{"40}
      \DeclareMathSymbol{\leqslant}{3}{AMSa}{"36}
      \DeclareMathSymbol{\geqslant}{3}{AMSa}{"3E}

      \let\leq=\leqslant \let\le=\leqslant
      \let\geq=\geqslant \let\ge=\geqslant
    \fi
  \fi
\fi % End of NFSS release 2

\ifCUPmtlplainloaded \else
  \ifAMStwofonts \else % If no AMS fonts
    \def\upi{\pi}
    \def\umu{\mu}
    \def\upartial{\partial}
  \fi
\fi

\title{Fragmentation of star-forming clouds enriched with the first dust}
\author[Raffaella Schneider, Kazuyuki Omukai, Akio Inoue \& Andrea Ferrara]
{Raffaella Schneider$^{1,2}$, Kazuyuki Omukai$^{3}$, Akio K. Inoue$^{4}$ \& Andrea Ferrara$^{5}$ \\
$^1$Centro Enrico Fermi, Via Panisperna 89/A, 00189 Roma, Italy\\
$^2$INAF/Osservatorio Astrofisico di Arcetri, Largo Enrico Fermi 5, 50125 Firenze, Italy \\
$^3$Division of Theoretical Astronomy, National Astronomical Observatory, Mitaka, Tokyo 181-8588, Japan \\
$^4$College of General Education, Osaka Sangyo University, 3-1-1 Nakagaito, Daito, Osaka 574-8530, Japan\\
$^5$SISSA/International School for Advanced Studies, Via Beirut 4, 34100 Trieste, Italy}
\date{}

\pagerange{\pageref{firstpage}--\pageref{lastpage}}
\pubyear{}

\begin{document}

\maketitle

\label{firstpage}

\begin{abstract}
The thermal and fragmentation properties of star-forming clouds have 
important consequences on the corresponding characteristic stellar mass.
The initial composition of the gas within these clouds is a record of
the nucleosynthetic products of previous stellar generations. In this
paper we present a model for the evolution of star-forming clouds 
enriched by metals and dust from the first supernovae, resulting from the explosions of 
metal-free progenitors with masses in the range 12 - 30 $\Msun$ and 
140 - 260 $\Msun$. Using a self-consistent approach, we
show that: (i) metals depleted onto dust grains play a fundamental role, 
enabling fragmentation to solar or sub-solar mass scales already
at metallicities $Z_{\rm cr} = 10^{-6} Z_{\odot}$; (ii) even at metallicities 
as high as $10^{-2} Z_{\odot}$, metals diffused in the gas-phase 
lead to fragment mass scales which are $\gsim 100 \Msun$; (iii) 
C atoms are strongly depleted onto amorphous carbon grains and CO molecules so 
that CII plays a minor role in gas cooling, leaving OI as the
main gas-phase cooling agent in low-metallicity clouds.   
These conclusions hold independently of the assumed supernova 
progenitors and suggest that the onset of low-mass star formation is 
conditioned to the presence of dust in the parent clouds.

\end{abstract}


\section{Introduction}

Supernovae represent the first cosmic polluters.  
On timescales shorter than 50 Myr, these stellar explosions seed 
the cosmic gas with the first metals. Theoretical models suggest that
a significant fraction of these heavy elements could be in the form
of dust grains (Kozasa, Hasegawa \& Nomoto 1989; Todini \& Ferrara
2001; Nozawa \etal 2003; Schneider, Ferrara \& Salvaterra 2004). 
Indeed, following the explosion, conditions are met in the expanding 
SN ejecta for the condensation of gas-phase
metals in grains of different size and species. In particular, 
the fraction of metals depleted onto dust, 
$f_{\rm dep} = M_{\rm dust}/M_{\rm met}$, varies 
between 0.3 and 0.7 for pair-instability supernovae (PISN) with 
progenitor masses in the range $140 - 260\Msun$ (Schneider \etal 2004), 
and between 0.24 and 1 for Type-II supernovae (SNII)
with $12 - 30 \Msun$ progenitors (Todini \& Ferrara 2001). 

Efficient dust production on short timescales in supernova ejecta provides
a straightforward interpretation of the large dust masses ($M_{\rm dust} \sim 10^9 \Msun$)
inferred from mm and submm observations of quasars (QSOs) with redshifts $5.7 < z <6.42$,
(Bertoldi \etal 2003: Priddey \etal 2003). 
Dust condensation in the atmospheres of AGB stars requires timescales of 1 - 2 Gyr (Morgan
\& Edmunds 2003), too long to account for the presence of large dust masses at 
redshifts $z > 5$, when the age of the Universe is $ < 1.2 $~Gyr. Further support to a SN
origin for dust at high redshift has recently come from near-IR observations of a $z=6.22$
QSO (Maiolino \etal 2004). The extinction properties of the observed spectrum cannot
be reproduced assuming a Small Magellanic Cloud dust extinction curve 
(which generally applies to $z < 5$ systems) and can, instead, be interpreted with
models for dust produced in SNII ejecta. 
Still, direct IR observations of SN remnants (Cassiopea A, Crab, Kepler) have 
detected only thermal emission from a hot dust component (temperatures $50 - 80$~K)
with associated dust masses of only $0.003 - 0.07 \Msun$ (Hines \etal 2004; 
Green \etal 2004). A larger
dust component could be at lower temperatures ($< 30$~K), and thus not visible at
IR wavelengths. Submm observations of some of these systems have, so far, yielded only
upper limits, with estimated masses $< 0.2 - 1.5 \Msun$ (Morgan \etal 2003; Krause \etal 2004; 
Wilson \& Batrla 2005). 

Early metal enrichment of the cosmic gas has important implications
for the formation of the first low-mass and long-lived stars. 
In the last few years, results of sophisticated numerical simulations 
(Abel, Brian \& Norman 2002; Bromm, Coppi \& Larson 2002) 
and semi-analytical models (Omukai \& Nishi 1998; Omukai 2000; 
Nakamura \& Umemura 2002) have shown that, due to the primordial
composition of the gas, the first stars were 
predominantly very massive, with masses of 
$100-1000~\Msun$. The presence of metals favors cooling
and fragmentation of prestellar-gas clouds, enabling the
formation of solar and sub-solar mass stars when the
gas initial metallicity reaches a critical value $Z_{\rm cr}$.
Despite its importance, theoretical studies have only
constrained this fundamental parameter to within a few orders
of magnitudes, $10^{-6} Z_{\odot} \leq Z_{\rm cr} \leq 5 \times 10^{-3} Z_{\odot}$,
with the lowest limit implied by models which include the effect
of dust cooling (Schneider, Ferrara, Natarajan \& Omukai 2002; Schneider \etal 2003; Omukai, Tsuribe, Schneider 
\& Ferrara 2005) and the highest limit suggested by studies where
metal line cooling is considered to be the main
driver of the transition (Bromm, Ferrara, Coppi \& Larson 1999). 

All these theoretical studies relies on the assumption that the 
relative metal abundances and dust grains properties which 
characterize the initial composition of pre-stellar gas clouds 
are the same as those observed in the local 
interstellar medium, simply rescaled to lower values of metallicity. 
However, the composition of the cosmic gas enriched by nucleosynthetic 
products of the first stellar generation through supernova explosions
is likely to be different. To overcome this basic limitation, 
Bromm \& Loeb (2003) have derived individual
critical abundances of C and O but have limited their analysis to 
CII and OI fine-structure line cooling during the relatively low-density 
evolution of collapsing pre-stellar clouds.

In this paper, we follow a different approach and study the 
detailed thermal and fragmentation history of collapsing clouds 
enriched by the first supernova explosions. We use the model developed
by Omukai (2000), and recently improved in Omukai \etal (2005) to
carry out the first self-consistent study which explicitly accounts for 
the relative metal abundances and dust properties predicted by supernova models. 
We follow the fate of collapsing clouds enriched by these nucleosynthetic products 
to investigate whether preferred mass scales can be identified, 
and eventually linked to characteristic masses of second generation
stars.

The initial composition of collapsing gas clouds must be inferred by results of stellar and
supernova models of metal-free stars. This relies on specific assumptions on the initial 
mass function (IMF) of Population III stars. Unfortunately, both the mass range and the
shape of the IMF are still hampered from our limited understanding of the accretion physics and 
protostellar feedback effects (Omukai \& Palla 2003; Bromm \& Loeb 2004; Tan \& McKee 2004; see also Bromm \& Larson 2004 
for a thorough review). Given these uncertainties, we consider Population III stars forming both in the 
SNII and PISN progenitor mass ranges, taking 22 $\Msun$ and 195 $\Msun$ stars as representative cases
for the two classes.

The paper is organized as follows: in section~2 we describe the characteristics of the model, 
with particular attention on those aspects which have been specifically introduced to account for 
the properties of supernova dust grains. In section~3 we present the results of the analysis, and
in section~4 we summarize the main conclusions and discuss their implications.

\section{The model}

The thermal and chemical evolution of star forming gas clouds is studied
using the model developed by Omukai (2000) and recently improved in 
Omukai \etal (2005). The collapsing gas clouds are described by a one-zone
approach where all physical quantities are evaluated at the center as a 
function of the central density of hydrogen nuclei, $n_{\rm H}$. The temperature
evolution is computed by solving the energy equation

\begin{equation}
\frac{de}{dt}=-p  \,\frac{d}{dt} \,\frac{1}{\rho} - \Lambda_{\rm net}
\label{eq:energy}
\end{equation}     
where the pressure, $p$,  and the specific thermal energy, $e$, are 
given by,

\be
p = \frac{\rho  \,k  \,T}{\mu  \,m_{\rm H}}
\ee

\be
e=\frac{1}{\gamma-1} \frac{k \,T}{\mu  \,m_{\rm H}}
\ee
\noindent
and $\rho$ is the central density, $T$ is the temperature, $\gamma$ is the adiabatic exponent, 
$\mu$ is the mean molecular weight and $m_{\rm H}$ is the mass of hydrogen nuclei. 
The terms on the right-hand side of the energy
equation are the compressional heating rate,

\be
\frac{d\rho}{dt} = \frac{\rho}{t_{\rm ff}}  \qquad \mbox{with} \qquad t_{\rm ff} = \sqrt{\frac{3 \pi}{32 G \rho}},
\ee

\noindent
and the net cooling rate, $\Lambda_{\rm net}$, which consists
of three components,

\be
\Lambda_{\rm net} = \Lambda_{\rm line} + \Lambda_{\rm cont} + \Lambda_{\rm chem}.
\ee
\noindent 
The first component, $\Lambda_{\rm line}$, represents the cooling rate due to the emission of line radiation,
which includes molecular line emission of H$_2$, HD, OH and H$_2$O, and atomic fine-structure line emission 
of CI, CII and OI. Following Omukai \etal (2005), H$_2$ collisional transition rates and HD parameters are
taken from Galli \& Palla (1998). The second component, $\Lambda_{\rm cont}$, represents the cooling rate due to
the emission of continuum radiation, which includes thermal emission by dust grains and H$_2$ collision-induced 
emission. The last term, $\Lambda_{\rm chem}$, indicates the cooling/heating rate due to chemical reactions.
When the gas cloud is optically thick to a specific cooling radiation, the cooling rate is correspondingly 
reduced by multiplying by the photon escape probability. Unless otherwise stated, the treatment of these processes
are the same as in Omukai (2000), to which we refer the interested reader for further details. 

In the present analysis, we relax the assumptions that gas-phase elemental abundances and dust grain 
properties are the same as those observed in local interstellar clouds. Instead, we compute the relative 
abundances of gas-phase metals and dust grains applying the model of Todini \& Ferrara (2001) for metal-free 
Type-II supernovae (SNII) and of Schneider, Ferrara \& Salvaterra (2003) for pair-instability supernovae (PISN). 
The properties of dust (relative dust compounds and grain size distributions) affect two processes which 
contribute to the thermal evolution of gas clouds: (i) the cooling rate due to dust thermal emission and 
(ii) the H$_2$ formation rate on dust grains. The description of these processes has been modified
with respect to the treatment in Omukai (2000) and Omukai \etal (2005) to properly take into account 
the properties of dust grains produced in the first supernova explosions.    

\subsection{Dust cooling}
\label{sec:dustcool}
The contribution of dust grains to gas cooling is determined by the balance between 
dust thermal emission and heating due to collisions with gas particles,

\begin{equation}
4 \, \sigma  \, T_{\rm gr}^4  \,\kappa_{\rm P}   \,\beta_{\rm esc}  \, \rho_{\rm gr} = \,H_{\rm gr}
\label{eq:tgrain}
\end{equation}

\noindent
where $\sigma$ is the Stefan-Boltzmann constant, $T_{\rm gr}$ is the dust temperature, 
$\kappa_{\rm P}$ is the Planck mean opacity of dust grains per 
unit dust mass, $\beta_{\rm esc}$ is the photon escape probability, and the dust mass density 
$\rho_{\rm gr}$ can be expressed as $\rho \, {\cal D}$, with ${\cal D}$ indicating the dust-to-gas ratio.
Following Omukai (2000) the photon escape probability can be written as,

\be
\beta_{\rm esc} = {\rm min}(1,\tau^{-2}) \qquad \mbox{with} \qquad \tau = \kappa_{\rm P} \rho \lambda_J,
\ee

\noindent
where the optical depth is estimated across one local Jeans length (roughly the
size of the central core region in the Penston-Larson similarity solution). 
The Planck mean opacity per unit dust mass is,

\begin{equation}
{\kappa}_{\rm P} = \frac{ \pi}{\sigma T_{\rm gr}^4} \int_0^{\infty} B_{\nu}(T_{\rm gr}) \kappa(\nu) d\nu
\label{eq:kplanck}
\end{equation}

\noindent
where $\kappa(\nu)$ is the absorption coefficient for the frequency $\nu$ 
per unit dust mass and $B_{\nu}(T_{\rm gr})$ is the Planck brightness.

The right-hand side of the eq.~(\ref{eq:tgrain}) is the collisional heating (cooling) rate per
unit volume (erg s$^{-1}$ cm$^{-3}$) of 
dust (gas) particles and can be written as (Hollenbach \& McKee 1979),

\begin{equation}
H_{\rm gr} = \frac{n_{\rm gr}(2 \, k  \,T - 2  \,k  \,T_{\rm gr})}{t_{\rm coll}}
\label{eq:heat}
\end{equation}  
\noindent
where $t_{\rm coll}^{-1} = n_{\rm H}\,\sigma_{\rm gr} \,\bar{v}_{\rm H}\,f$ is the average time between two successive
collisions, $n_{\rm gr}$ and $\sigma_{\rm gr}$ are the grain number density and cross section, $\bar{v}_{\rm H}$ is 
the average speed of hydrogen nuclei and $f = \bar{n v}/n_{\rm H}\bar{v}_{\rm H}$ measures the contribution of 
other species.
We assume that the gas is fully molecular and follows a Maxwellian distribution so that,
\be
\bar{v}_{\rm H} = \left(\frac{8 \,k \,T}{\pi \,m_{\rm H}}\right)^{1/2}, 
\label{eq:vel}
\ee
\noindent
and $f = 0.3536 + 0.5 \, y_{\rm He}$, where $y_{\rm He} = n_{\rm He}/n_{\rm H} = 0.083$ 
(which corresponds to a He mass fraction of Y=0.25) and we have neglected the contribution of
heavy elements. These are good approximations in the density range where collisional dust heating is relevant 
($n_{\rm H} > 10^{12-13} \mbox{cm}^{-3}$) and for the gas metallicities that we are interested to 
($Z = Z_{\rm cr}$).

\noindent
If we further write,
\begin{equation} 
n_{\rm gr} \,\sigma_{\rm gr} = n_{\rm H} \,  m_{\rm H} (1 + 4 y_{\rm He}) \, S \, {\cal D} 
\label{eq:cross}
\end{equation}
\noindent
in eq.~(\ref{eq:heat}), where $S$ is the total grain cross section per unit mass of dust, eq.~(\ref{eq:tgrain}) appears
to be independent of the dust-to-gas ratio and to depend on the thermal states of gas and dust and on grain specific
properties (cross section and Planck mean opacity). 

%%%%%%%%%%%%%%%% TABLE 1 %%%%%%%%%%%%%%%%%%
\begin{table*}
\begin{center}
\caption{\footnotesize
Properties of dust grains produced in the ejecta of a Z=0, 22 $\Msun$ SNII and of a 195 $\Msun$ PISN. The table show various grain species (first column), their mass fraction relative to the total dust mass produced (second and fourth columns), and their geometrical cross-sections (third and fifth columns). References for the adopted grain optical properties are given in the last column. In some cases, data was linearly extrapolated on the log-log plot to extend the original wavelength coverage to longer wavelengths.}  
\begin{tabular}{l|c|c|c|c|l}\hline
  Material     & $f_{i}$         & $S_i$          & $f_{i}$      & $S_i$       & Optical properties  \\ \hline

               & 195 $\Msun$     & 195 $\Msun$    &  22 $\Msun$     & 22 $\Msun$     &                     \\ \hline
 Al$_2$O$_3$   & 5.82\,10$^{-4}$ & 4.82\,10$^{5}$ & 5.82\,10$^{-3}$ & 1.36\,10$^{6}$ & ISAS sample from Koike \etal (1995)\\ 
 Fe$_3$O$_4$   & 1.33\,10$^{-2}$ & 7.00\,10$^{4}$ & 2.88\,10$^{-1}$ & 2.58\,10$^{5}$ & Mukai (1989) \\
 MgSiO$_3$     &   /             &      /         & 5.24\,10$^{-2}$  & 3.22\,10$^{6}$ & Semenov \etal (2003) \\
 Mg$_2$SiO$_4$ & 0.217           & 6.30\,10$^{4}$ & 3.43\,10$^{-1}$ & 1.40\,10$^{6}$ & Semenov \etal (2003) \\
 SiO$_2$       & 0.705           & 5.58\,10$^{4}$ &    /            &     /          & Henning \& Mutschke (1997); Philipp (1985) \\ 
 AC (amorphous carbon)            & 6.41\,10$^{-2}$ & 8.06\,10$^{4}$ & 3.11\,10$^{-1}$ & 8.50\,10$^{4}$ & ACAR from Zubko \etal (1996) \\   
\hline
\end{tabular}
\end{center}
\label{tbl:grain}
\end{table*}
%%%%%%%%%%%%%%%%%%%%%%%%%%%%%%%%%%%%%%%%%%%

Given the grain size distribution functions per unit dust mass (cm$^{-1}$ gr$^{-1}$), 
$n_{\rm gr, i}(a) =dn_{\rm gr, i}(a)/da$, and the mass fractional abundance, $f_i$, of each grain type $i$ 
predicted by the SN-dust formation models, the total grain cross section per unit dust mass (cm$^{2}$/gr) is,

\be
S = \Sigma_i \, f_i \, S_i  \qquad  \mbox{with}  \qquad  S_i =
\int_0^{\infty} \, n_{\rm gr, i}(a) \, \pi \, a^2 \, da. \nonumber
\label{eq:crossdust}
\ee

\begin{figure}
\center{{
\epsfig{figure=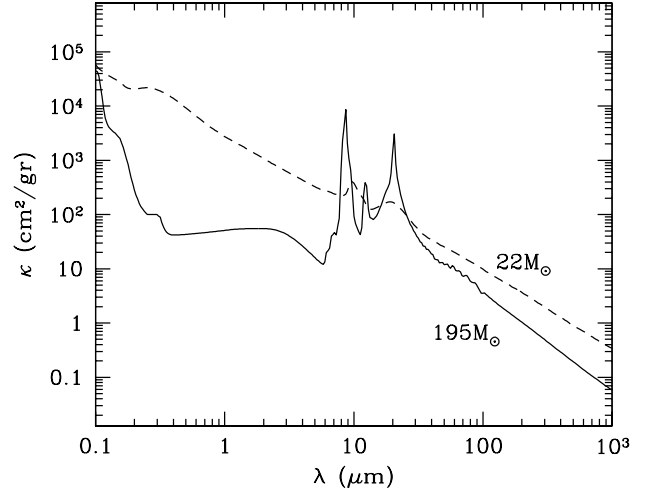,height=6.5cm }
}}
\caption{Absorption coefficients per unit dust mass assuming the properties 
(size distributions and relative masses of different grain species) of dust grains predicted to form in the
ejecta of a 195 $\Msun$ PISN (solid lines) and of a 22 $\Msun$ SN (dashed lines).}
\label{fig:absext}
\end{figure}

\noindent  
Similarly, the total absorption coefficient per unit dust mass (cm$^{2}$/gr) is,

\be
\kappa(\nu) = \Sigma_i \, f_i \, \kappa_{\nu, i} \, \, \, \,  \mbox{with}
\, \, \, \, 
\kappa_{\nu, i} = \int_0^{\infty} Q^i_{\nu}(a) \, n_{\rm gr, i}(a) \,
\pi \, a^2 \, da,  \nonumber
\label{eq:absdust}
\ee
\noindent
where $Q^i_{\nu}$ are the absorption (or extinction) cross section normalized to the geometrical cross-sections. 
For each dust species, the $Q$ parameters for any frequency and radius have been computed using the standard
Mie theory for spherical grains with grain optical constants collected from published data (see Table~1
for details). In Fig.~\ref{fig:absext} we show the total absorption coefficients per unit dust mass
assuming the properties of dust grains formed in the ejecta of a 195 $\Msun$ PISN 
(see Figs.~4 and ~5 in Schneider, Ferrara \& Salvaterra 2004) and of a Z=0, 22 $\Msun$ SN 
(see Figs.~5 and ~6 in Todini \& Ferrara 2001). For the PISN model, the absorption coefficient 
is dominated by SiO$_2$ grains in the 6 $\mu$m - 200 $\mu$m range, whereas above 200 $\mu$m amorphous carbon (AC) 
grains give a comparable contribution. Shortward of 6 $\mu$m, AC grains largely 
dominate down to $\sim 0.2 \mu$m, below which Mg$_2$SiO$_4$ grains produce the steepening of the curve. 
For the SNII model, the absorption coefficient is largely dominated by AC grains except
for the 0.8 $\mu$m - 30 $\mu$m range, where the Mg$_2$SiO$_4$ features appear.  

%%%%%%%%%%%%%%%%%%%%%%%%%%%%%%%%%%%%%%%%%%%%%%%%%%%%%%%
%%%%%%%%%%%%%%%%%%%%%%%%%%%%%%%%%%%%%%%%%%%%%%%%%%%%%%%

\begin{figure}
\center{{
\epsfig{figure=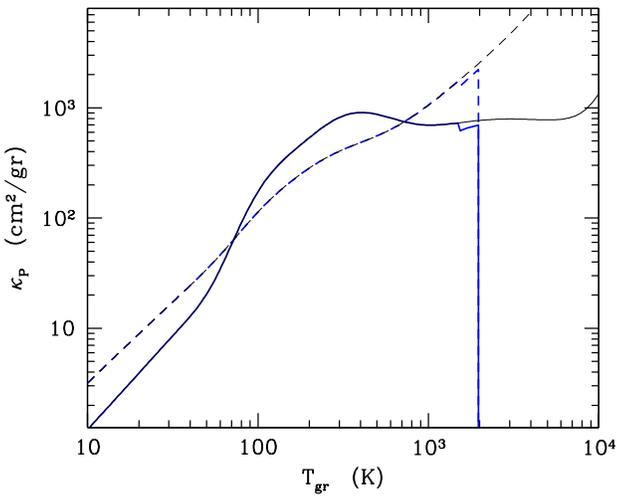,height=6.5cm }
}}
\caption{Planck mean opacities per unit dust mass for absorption for the 195 $\Msun$ PISN model (solid)
and for the 22 $\Msun$ SN case (dashed). Thin lines assume that the grains never sublimate (see text).}
\label{fig:kp}
\end{figure}

The resulting Planck mean opacities are shown in Fig.~\ref{fig:kp}. The solid
line shows the absorption Planck mean per unit
dust mass for the 195 $\Msun$ PISN model and the dashed line shows the
same result for the 22 $\Msun$ model. 
When computing eq.~(\ref{eq:kplanck}), it is important to
take into account the effect of dust sublimation: grains can be destroyed by heating through collisions with gas
particles if the equilibrium grain temperature $T_{\rm gr}$ exceeds a characteristic value $T^{i}_{\rm sub}$ which, 
for each grain species $i$, depends on the gas density through the relative partial pressures of the heavy elements
in the gas phase. Here we assume a sublimation temperature of $T^{AC}_{\rm sub} = 2000~$K for AC grains and of 1500~K
for other grain species (Laor \& Draine 1993; Preibisch \etal 1993) and ignore the density dependence
\footnote{Note that comparable values are found for the condensation temperatures in the 195 $\Msun$ model, 
namely 2100~K for AC grains, 1800~K for Al$_2$O$_3$, and 1470~K $<$ T $<$ 1560~K for the other grain species (see
Todini \& Ferrara 2001; Schneider \etal 2004).}.  
The thin lines in Fig.~\ref{fig:kp} show the results when sublimation
is not considered.  

%%%%%%%%%%%%%%%%%%%%%%%%%%%%%%%%%%%%%%%%%%%%%%%%%%%%%%%
%%%%%%%%%%%%%%%%%%%%%%%%%%%%%%%%%%%%%%%%%%%%%%%%%%%%%%%

\subsection{H2 formation on dust grains}

An H atom can be bound to the grain surface in two energetically different sites: a physisorbed site, when 
the attractive forces are due to mutually induced dipole moments in the electron shells of H and surface 
atoms (van der Waals interactions), and a chemisorbed site, when overlap between their respective 
wave-functions occur. Coming from the gas phase, the H atoms are first physisorbed and then either cross the
barrier between the physisorbed and chemisorbed sites (by moving perpendicularly to the surface), or go to 
another physisorbed site (by moving along the surface). The relevant parameters which regulate grains 
surface characteristics are therefore the desorption energies of physisorbed H, $E_{\rm HP}$, of chemisorbed
 H, $E_{\rm HC}$, and of the saddle point between a physisorbed and a
 chemisorbed site, $E_{\rm S}$, and can 
be derived by laboratory experiments for different grain species 
(see Cazeaux \& Spaans 2004 and references therein)\footnote{From
  Table~1 of Cazaux \& Spaans (2004), 
we take $E_{\rm HP} = 800$~K, $E_{\rm HC} = 3 \times 10^4$~K, and $E_{\rm S} = 250$~K for AC grains. 
For other grain species we assume the same parameters as those derived for olivine (Mg$_2$SiO$_4$), namely
$E_{\rm HP} = 650$~K, $E_{\rm HC} = 3 \times 10^4$~K, and $E_{\rm S} =
200$~K }.

At temperatures where the process of H$_2$ formation on grain surface
is relevant in collapsing gas clouds, the most efficient reaction is
the ``collision'' of two chemisorbed atoms, with efficiency (Cazeaux
\& Tielens 2002), 

\be
\epsilon_{\rm H_2} = \left (1 + \frac{\beta_{\rm HP}}{\alpha_{\rm PC}}\right)^{-1} 
\ee
\noindent
where we have neglected the correction factor at high temperatures
($T>300$~K) which reflects the evaporation of chemisorbed H. In the
above equation, $\beta_{\rm HP}$ is the evaporation rate of
physisorbed
H and $\alpha_{\rm PC}$ is the mobility to go from a physisorbed to a
chemisorbed site, and their ratio can be approximated as (Cazeaux
\& Tielens 2002),

\be
\frac{\beta_{\rm HP}}{\alpha_{\rm PC}} = \frac{1}{4} \left (1+\sqrt{\frac{E_{\rm HC} - E_{\rm S}}{E_{\rm HP} - E_{\rm S}}} \right)^{2} \exp{\left[-\frac{E_{\rm S}}{kT_{\rm gr}}\right]}.
\ee
\noindent
The resulting H$_2$ formation rate per unit volume (cm$^{-3}$ s$^{-1}$) on grain
surface can be expressed as,

\begin{eqnarray}
R_{\rm H_2}&=&\frac{1}{2} \, n({\rm H}) \, \bar{v}_{\rm H} \, n_{\rm
           gr} \, \sigma_{\rm gr} \, {\epsilon}_{\rm H_2} \, S_{\rm
           H}(T)  \nonumber \\
           &\equiv& k_{\rm gr} \, n({\rm H}) \, n_{\rm H}  
\label{eqn:H2rec}
\end{eqnarray}

\noindent
where $n({\rm H})$ is the number density of H atoms in the gas phase, $n_{\rm H}$
is the number density of H nuclei (protons),
and $S_{\rm H}(T)$ is the sticking coefficient of H atoms which 
depends both on dust and gas temperatures (Hollenbach \& McKee 1979),

\begin{eqnarray}
&S_{\rm H}(T) =   \\  
&\left[1 + 0.4\left(\frac{T + T_{\rm gr}}{100~{\rm
      K}}\right)^{0.5}+0.2\left(\frac{T}{100~{\rm K}}\right) 
+ 0.08 \left(\frac{T}{100~{\rm K}}\right)^2\right]^{-1}. \nonumber
\end{eqnarray}

\noindent
Substituting eqs.~(\ref{eq:vel}) and~(\ref{eq:cross}) in eq.~(\ref{eqn:H2rec}), the coefficient $k_{\rm gr}$ 
can be written as,
\be
k_{\rm gr} =  \left (\frac{2\,k\,T}{m_{\rm H} \pi} \right)^{1/2} \, S_{\rm H}(T) \,{\cal D} \, m_{\rm H}\, (1+4 y_{\rm He}) \, \Sigma_{i} f_i \, S_i \, \epsilon_{\rm H_2}^i,
\ee   
\noindent
where $\epsilon_{\rm H_2}^i$ is the H$_2$ recombination efficiency for
each grain species.

\section{Results}

\subsection{Thermal evolution of collapsing clouds}

Depending on the efficiency of metal diffusion and mixing, 
gas clouds in the vicinity of supernovae can be enriched 
to a wide range of metallicity values. If we assume that
gas-phase metals and dust grains are not selectively 
transported, we can take the initial composition
of pre-stellar gas clouds to reflect the final 
composition of SN ejecta. We then follow the thermal evolution 
of collapsing gas for a set of initial metallicities but relative metal 
and dust properties consistent with those predicted 
by the corresponding supernova model. 

%%%%%%%%%%%%%%%%%%%%%%%%%%%%%%%%%%%%%%%%%%%%%%%%%%%%%%%
\begin{figure*}
\center{{
\epsfig{figure=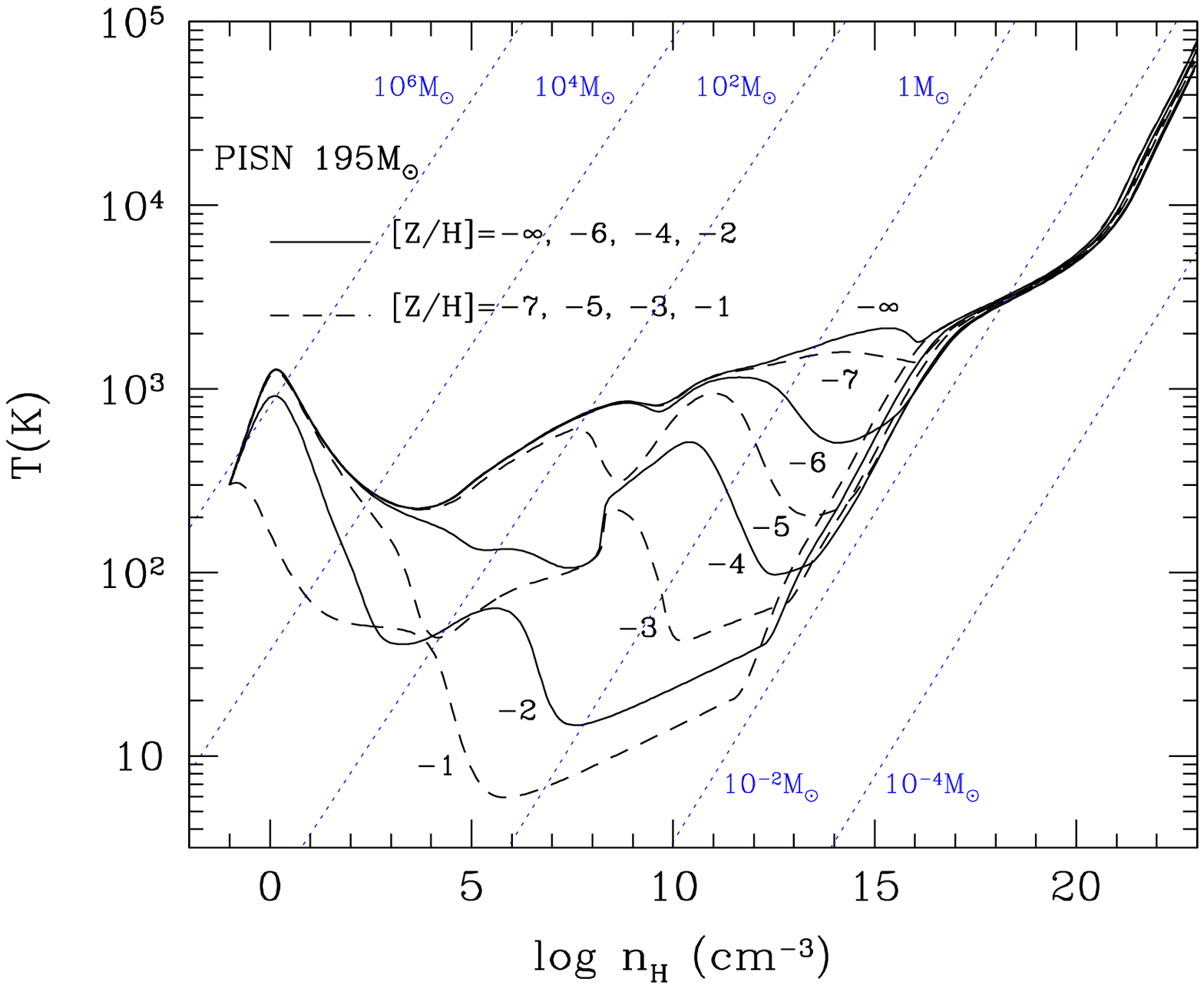,height=6.5cm}
\hspace{1.0cm}
\epsfig{figure=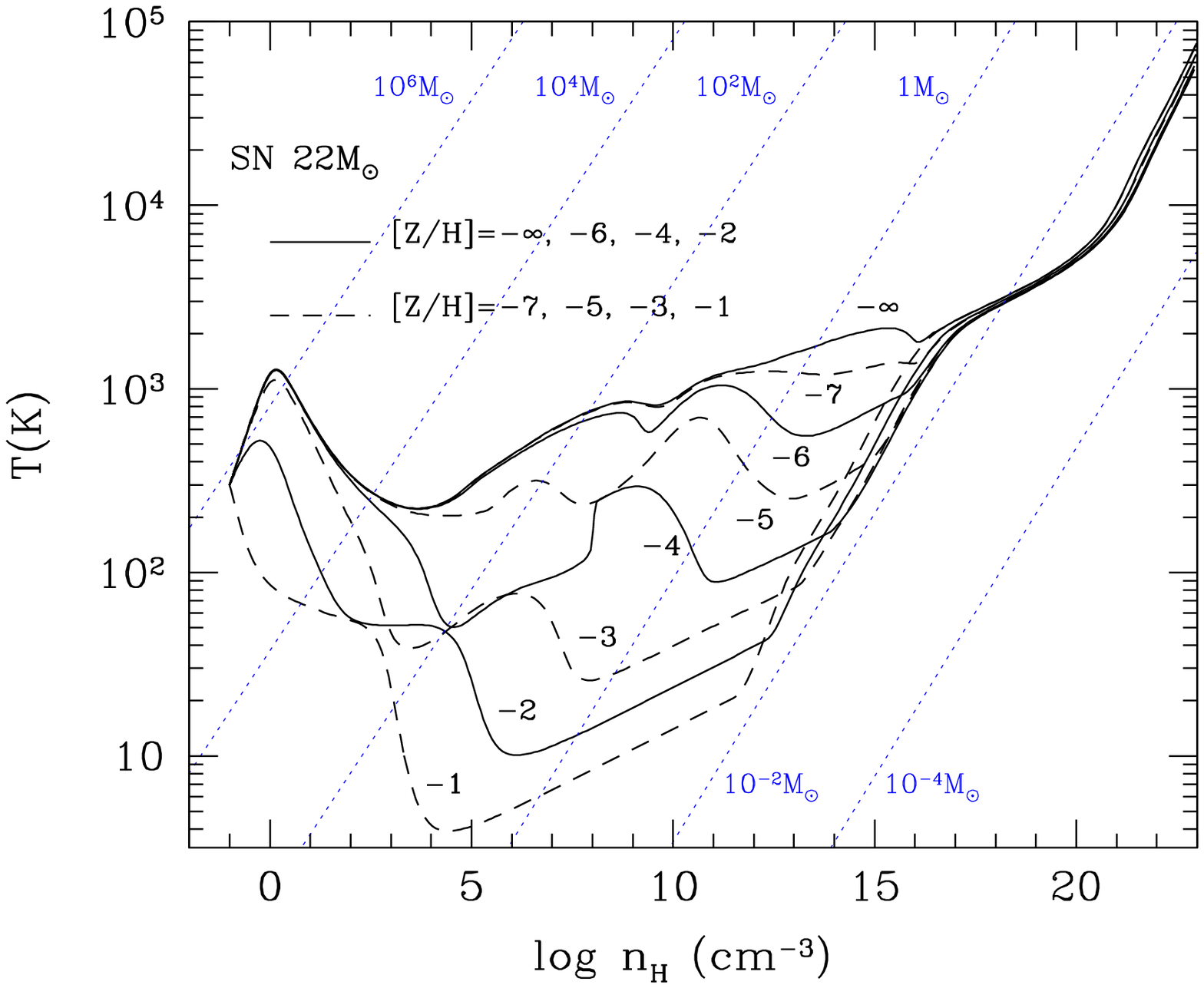,height=6.5cm}
}}
\caption{Thermal evolution of collapsing pre-stellar
gas clouds as a function of the central gas number density 
enriched to different metallicities by the
explosion of a PISN 195 $\Msun$ (left) and of a zero metallicity
22 SNII (right). Solid curves indicate total metallicities 
(gas-phase and solid dust grains) 
[Z/H]=Log$(Z/Z_{\odot}) = -\infty, -6, -4, -2$
and dashed curves to [Z/H] = -7, -5, -3, -1. 
The diagonal dotted curves identify values of
the Jeans mass corresponding to given thermal states.}
\label{fig:nT}
\end{figure*}
%%%%%%%%%%%%%%%%%%%%%%%%%%%%%%%%%%%%%%%%%%%%%%%%%%%%%%%

The results are shown in Fig.~\ref{fig:nT} for 
gas clouds with initial metallicities in the range 
$0 \le Z/Z_{\odot} \le 0.1$ enriched by the products of
a 195 $\Msun$ PISN (left panel) and of a 22 $\Msun$ 
metal-free SNII (right panel).
For these specific supernova models, dust depletion factors 
are predicted to be $f_{\rm dep}= 0.65$ and 0.24, respectively. 
Consistent with previous findings (Omukai 2000; Schneider \etal
2002, 2003; Omukai \etal 2005), at relatively low densities 
($n_{\rm H} \lsim 10^5$cm$^{-3}$), thermal evolution is
dominated by line-cooling of H$_2$ for $Z/Z_{\odot} \lsim 10^{-5}$,
of HD for $10^{-4} \lsim Z/Z_{\odot} \lsim 10^{-3}$, and, for $Z/Z_{\odot} \gsim 10^{-2}$, 
of O fine structure lines. Once the relevant cooling agent
reaches the NLTE - LTE transition or the gas becomes optically 
thick to the dominant cooling radiation, line-cooling becomes inefficient and gravitational 
contraction leads to a temperature increase. 

Note that, contrary to what commonly assumed, the above results suggest that 
C makes a negligible contribution to fine-structure line cooling, being
strongly depleted onto AC grains and CO molecules for most SN models (Todini \& Ferrara 2001;
Schneider \etal 2004).

The impact of dust-cooling on the thermal evolution starts to be apparent for initial
metallicities $Z/Z_{\odot} \gsim 10^{-6}$ ($Z \ge Z_{\rm cr}$) as a temperature dip
at high densities ($n_{\rm H} \gsim 10^{13}$cm$^{-3}$), which progressively shifts
to lower densities for increasing total metallicity\footnote{If a larger amount of dust is present
because the total metallicity of the gas is higher, then the corresponding cooling 
rate starts to be effective earlier in the evolution of collapsing gas clouds and
it is not confined to the highest density regime.}, until, for $Z/Z_{\odot} \gsim 10^{-2}$,
the temperature dips due to line-cooling and dust-cooling merge and
it is no-longer possible to separate the two contributions.  

\subsection{The impact of SN dust}

It is interesting to compare the results of the self-consistent model 
for thermal evolution of collapsing clouds with previous studies, such
as in Omukai \etal (2005), based on different assumptions on dust and 
metal properties. Comparing the tracks shown in Fig.~\ref{fig:nT} with 
the results in Fig.~1 of Omukai \etal (2005), we see that the temperature
dip at high density ($n_{\rm H} \sim 10^{13}$~cm$^{-3}$) due to dust-cooling
which in the present analysis is already apparent in the $Z = 10^{-6} Z_{\odot}$
track, starts to be evident only at higher metallicities ($Z \ge 10^{-5} Z_{\odot}$)
in Omukai \etal (2005). This result reflects a higher cooling efficiency of 
SN-dust with respect to present-day interstellar dust, partly due to larger 
grain cross-section and absorption coefficient per unit dust mass 
(eqs.\ref{eq:crossdust}-\ref{eq:absdust}). Indeed, due the limited time available
for grain-growth, SN-dust grains are smaller than present-day interstellar
grains and the ratio of area to mass tends to be larger for smaller grains
(Todini \& Ferrara 2001; Nozawa \etal 2003; Schneider \etal 2004). 
A major difference is represented by the condensed C species adopted: in SN-dust
models, C atoms condense in AC grains which are characterized by a 
high sublimation temperature (see section~\ref{sec:dustcool}) and survive to 
the high-temperature environment of low-metallicity clouds. Conversely, 
following Pollack \etal (1994) all previous models assumed that C atoms 
condense in organics and therefore sublimate at lower temperatures, $\lsim 600$~K
(Omukai 2000; Schneider \etal 2003; Omukai \etal 2005).

The larger grain cross-section per unit mass of SN-dust 
enhances the H$_2$ formation rate. However, this has only
a minor effect on the thermal evolutionary tracks at the
lowest metallicities, being relevant only for 
$Z \gsim 10^{-4} Z_{\odot}$.

\subsection{Fragmentation properties}
Our main interest here is to investigate the 
fragmentation properties of clouds enriched by primordial supernovae, with the aim
of understanding whether a preferred mass scale can be identified and eventually related
to the characteristic mass of second-generation stars. 
Indeed, numerical simulations of primordial star formation have shown that
in primordial environments, where magnetic fields, turbulence and rotation
are likely to play a minor role in the dynamics, a preferred mass scale 
can be related to the Jeans mass at the end of the fragmentation process (see
Bromm \& Larson 2004 and references therein). More generally, Larson (2005)
suggests that the low-mass, Salpeter-like IMF, which characterizes contemporary
 star formation, is determined largely by thermal physics and fragmentation in
pre-stellar clouds.

%%%%%%%%%%%%%%%%%%%%%%%%%%%%%%%%%%%%%%%%%%%%%%%%%%%%%%%
\begin{figure*}
\center{{
\epsfig{figure=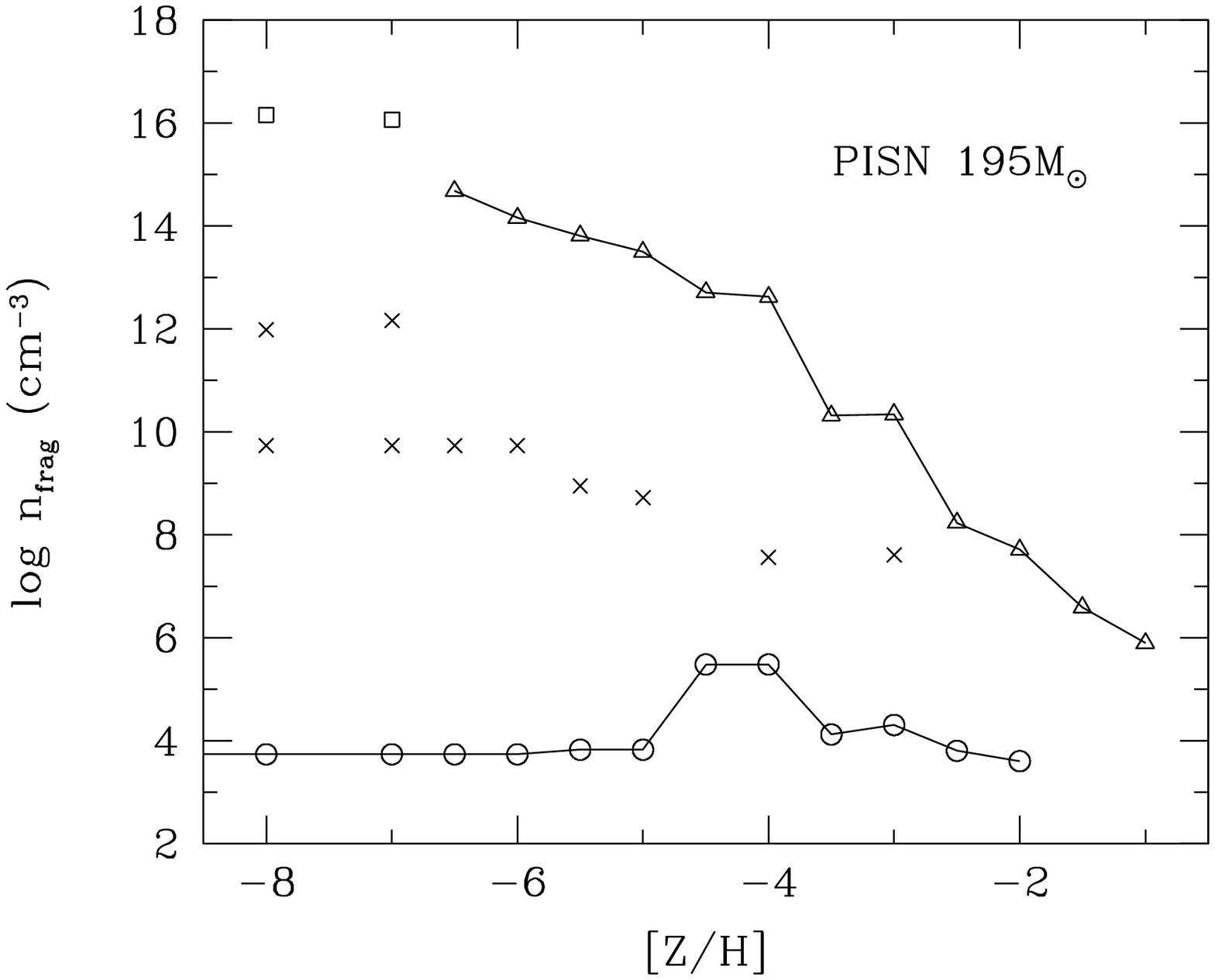,height=6.5cm}
\hspace{1.0cm}
\epsfig{figure=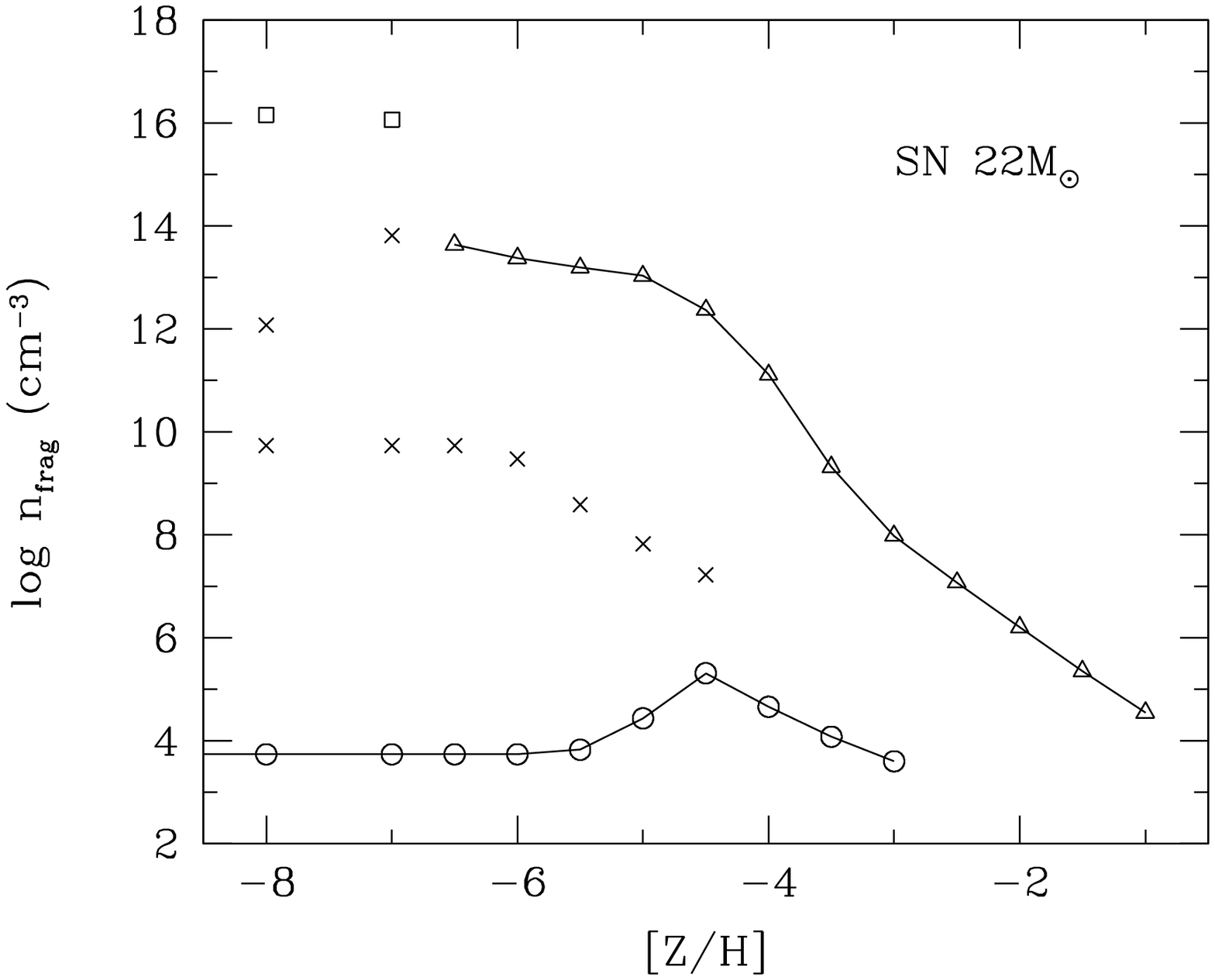,height=6.5cm}
}}
\caption{Gas number density where $\gamma = 1$ as a function of
total metallicity (gas-phase metals and solid grains), for pre-stellar
clouds enriched by the explosion of a 195 $\Msun$ PISN (left panel) and 
a 22 $\Msun$ SNII (right panel). Points indicated with circles (triangles)
represent fragmentation induced by line (dust) cooling. Crosses show the
effect of 3-body H$_2$ formation and H$_2$O cooling and, at
the lowest metallicities, squares indicate fragmentation due to H$_2$ 
collision-induced emission (see text).}
\label{fig:ngamma}
\end{figure*}
%%%%%%%%%%%%%%%%%%%%%%%%%%%%%%%%%%%%%%%%%%%%%%%%%%%%%%%

Fragmentation occurs efficiently when the adiabatic index 
$\gamma = {\rm dlogp/dlog\rho} < 1$,
and almost stops when isothermality breaks ($\gamma > 1$) as also shown
by the simulations of Li, Klesser \& MacLow (2003). Thus, consistent
with Schneider \etal (2002), (2003) we can adopt the condition $\gamma=1$ to
identify the preferred mass scale of the fragments (Inutsuka \& Miyama 1997; 
Jappsen \etal 2005)

In Fig.~\ref{fig:ngamma} we plot the pre-stellar
gas number density where this condition is met as a function of the initial total metallicity,
for pre-stellar clouds enriched by the explosion of a $195 \Msun$ PISN (left panel) and zero
metallicity 22 $\Msun$ SNII (right panel). The ``$\gamma=1$'' points can be classified into
three main branches: the lowest-density points, shown as open circles, indicate fragmentation
epochs which follow the temperature dip induced by line-cooling (hereafter line-induced fragmentation);
the highest-density points represent fragmentation following cooling due to 
H$_2$ collision-induced emission at the lowermost metallicities (squares) and due to dust thermal
emission (hereafter dust-induced fragmentation) at higher metallicities (triangles); for completeness, 
we also show a third branch of $\gamma =1$ points (crosses) which corresponds to gas number densities 
$n_{\rm frag} \sim 10^{10}$ cm$^{-3}$, and which is associated to cooling by 3-body H$_2$ formation and H$_2$O. As it
can be inferred from Fig.~\ref{fig:nT}, the cooling rate at these intermediate densities leads to temperature
decrements which tend to be short-lived and shallow and are unlikely to induce fragmentation.
The local maxima of the line-induced fragmentation branch result from the competition
between HD cooling (which starts to be effective at $Z \gsim 10^{-5} Z_{\odot}$) and heating due to 
H$_2$ formation on grain surface which, for $Z \gsim 10^{-4} Z_{\odot}$ prevents the gas from reaching
the NLTE-LTE transition for HD at densities $n \sim 10^5$cm$^{-3}$. 

It is important to stress that the presence of a temperature dip in the
thermal evolutionary track as well as the condition that $\gamma<1$ imply
only the possibility of fragmentation. For clouds of solar
metallicity, recent three-dimensional hydrodynamical simulations
show that the single temperature minimum present at a density comparable
to that of observed cloud cores plays an important role in determining
the peak mass of the IMF (Jappsen \etal 2005). In contrast, the second
minimum in the thermal evolutionary track due to dust-cooling occurs
at a density which is several order of magnitudes higher, deep in the
interior of the collapsing cloud. Therefore, it is not clear whether
the same argument may be applied as well as the fraction of mass which
may be affected. 

Dust-induced fragmentation will occur if very flattened or elongated structures 
form in the dense central region, and if the fragments formed do not 
subsequently merge. Indeed, following the first line-induced fragmentation, 
pre-stellar gas clouds experience runaway collapse, during which density 
perturbations might be erased by pressure forces. If so, with 
little elongation of the densest core, dust-induced fragmentation at higher 
densities would be prohibited. By means of a semi-analytic model for the 
linear evolution of non-spherical deformations, Omukai \etal
(2005) discuss the fragmentation histories of collapsing pre-stellar 
clouds and confirm that dust-induced
fragmentation does occur even for metallicities as low as $Z=10^{-6}Z_{\odot}$. 
The mass fraction in low-mass fragments is initially small
but becomes dominant for $Z \gsim 10^{-5} Z_{\odot}$ and continues to 
grow with increasing $Z$.
More recently, Tsuribe \& Omukai (2006) used three-dimensional
 hydrodynamical simulations to follow the evolution of 
low-metallicity cores during the dust-cooling phase and discuss the
conditions for fragmentation into low-mass clumps. Their results 
confirm that the previous analysis of Omukai \etal (2005) was 
broadly correct: fragmentation does occur in the
dust-cooling phase and sub-solar mass fragments are indeed produced.

\subsection{Characteristic mass}

\begin{figure*}
\center{{
\epsfig{figure=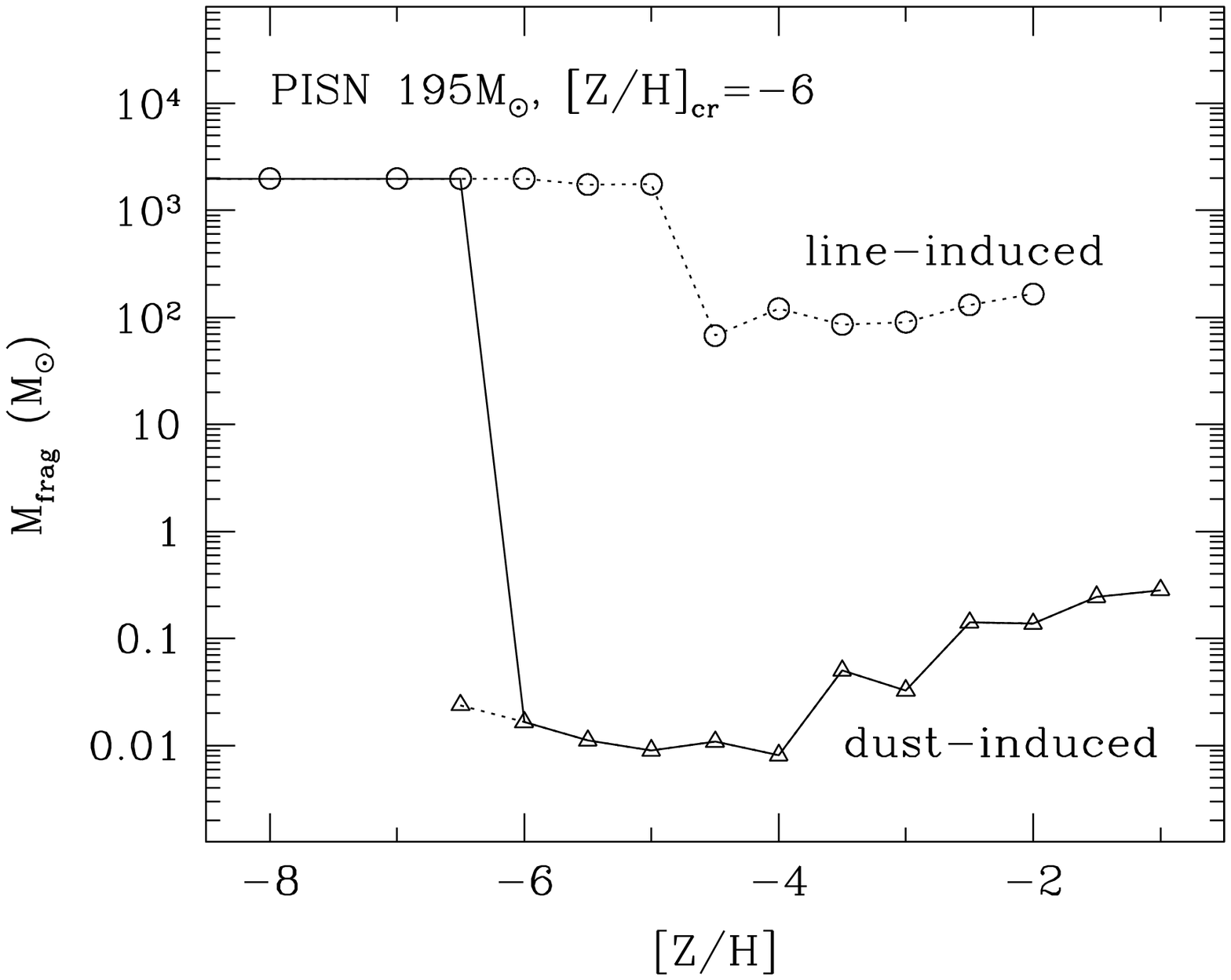,height=6.5cm}
\hspace{1.0cm}
\epsfig{figure=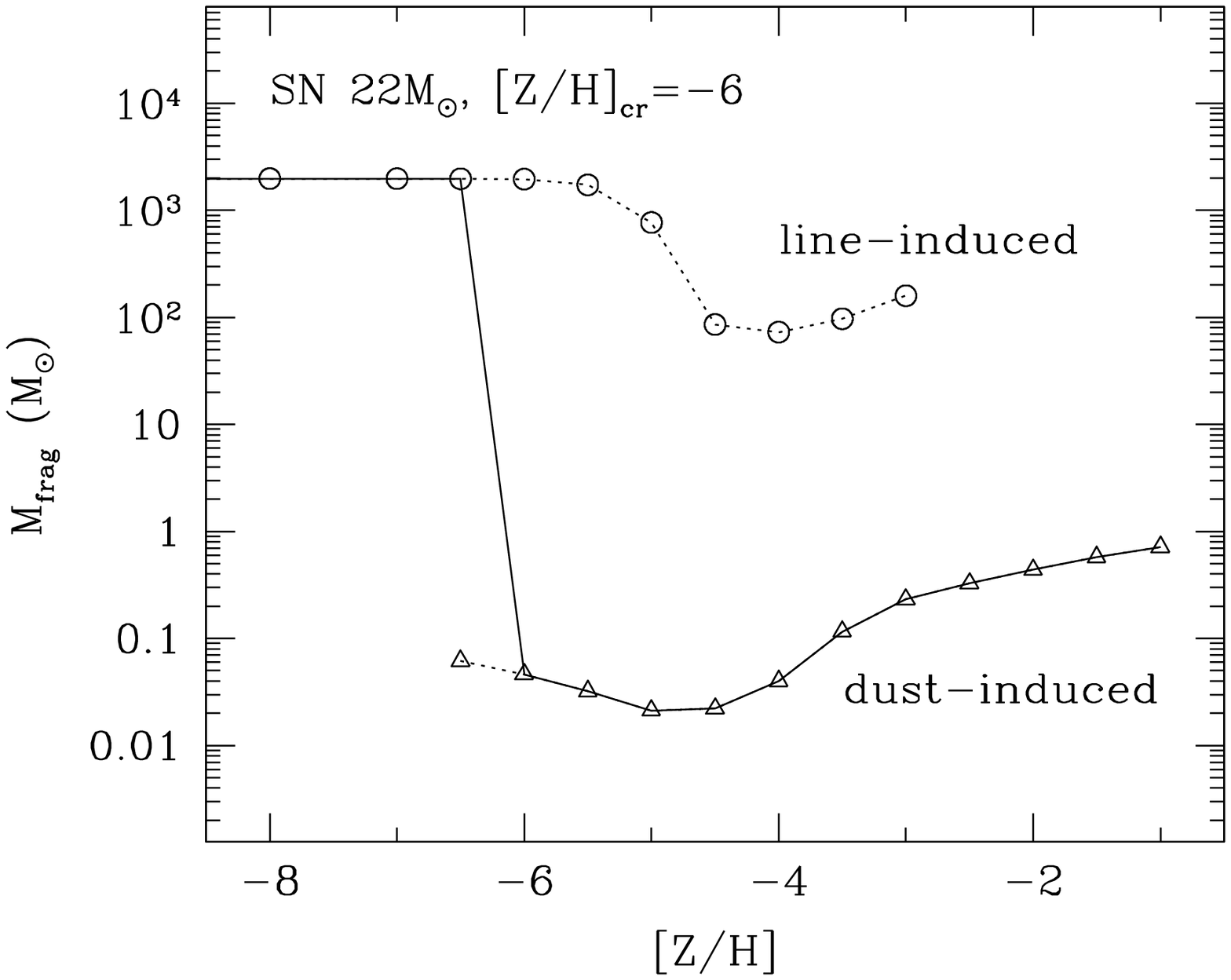,height=6.5cm}
}}
\caption{Thermal Jeans mass corresponding to line-induced fragmentation
(circles in Fig.~\ref{fig:ngamma}) and dust-induced fragmentation (triangles
in Fig.~\ref{fig:ngamma}) as a function of total metallicity, for pre-stellar
clouds enriched by the explosion of a 195 $\Msun$ PISN (left panel) and 
a 22 $\Msun$ SNII (right panel). The solid line indicates the transition
in characteristic fragment masses when the dust-induced fragmentation mode 
is activated at a critical metallicity $Z_{\rm cr} = 10^{-6} Z_{\odot}$.}
\label{fig:Mj}
\end{figure*}

In the present context, where we assume that thermal pressure is the main force opposing gravity, 
the characteristic mass of each fragment is given by the Jeans mass at fragmentation epoch,
\[
M_{\rm frag} = M_{J}(n_{\rm frag}, T_{\rm frag}) \propto \rho \lambda_J^3 
\propto T_{\rm frag}^{3/2} n_{\rm frag}^{-1/2}. 
\]
Fig.~\ref{fig:Mj} shows the Jeans masses associated to the lower and upper density branches represented
in Fig.~\ref{fig:ngamma}, hence to line-induced and dust-induced fragmentation. Since line-induced fragmentation 
occurs at relatively low densities, the corresponding fragment masses are large. In particular, at the lowest
metallicities we find that the characteristic fragmentation mass scale, induced by H$_2$ line cooling, 
is $M_{\rm frag} \sim 10^3 \Msun$, consistent with the results of numerical models. Still, even for total 
metallicities in the range $10^{-5} \le Z/Z_{\odot} \le 10^{-2}$, metal-line cooling leads to fragment masses 
$M_{\rm frag} \ge 100 \Msun$. 

Conversely, dust-induced fragmentation leads to solar or sub-solar mass fragments, with 
$0.01 \Msun \le M_{\rm frag} \le 1 \Msun$. Indeed, this fragmentation mode occurs at 
much higher densities (see Fig.~\ref{fig:ngamma}) where the associated Jeans masses 
are correspondingly lower. The characteristic fragment mass increases for increasing metallicity
reflecting the opposite trend in the dust-induced fragmentation density shown in Fig.~\ref{fig:ngamma}.
For both supernova models, this fragmentation mode is active already at very low metallicities
and the solid line illustrates the transition in fragmentation mass scales for $Z_{\rm cr} = 10^{-6} Z_{\odot}$.

The importance of dust for the formation of low-mass stars was
first emphasized by the pioneering work of Low \& Lynden-Bell (1976), where 
they also recognized that dust mantains its cooling efficiency down to 
a reduction by mass of $10^{-5}$ of the present-day value. 

The results that we find suggest that in the absence of dust 
pre-stellar clouds enriched by primordial supernovae
are not able to form solar or sub-solar mass fragments.

%%%%%%%%%%%%%%%%%%%%%%%%%%%%%%%%%%%%%%%%%%%%%%%%%%%%%%%
%%%%%%%%%%%%%%%%%%%%%%%%%%%%%%%%%%%%%%%%%%%%%%%%%%%%%%%

\section{Discussion}

In this paper we have presented the first self-consistent model of 
thermal evolution of pre-stellar gas clouds enriched with metals and dust
by primordial supernovae. We have focused on two representative 
cases: a 195 $\Msun$ pair-instability supernova and a
22 $\Msun$ TypeII supernova. Using the one-zone semi-analytic model initially 
developed by Omukai (2000), and recently improved in Omukai \etal (2005), 
we have followed the thermal and fragmentation history of pre-stellar clouds 
enriched by the products of single primordial supernovae. 
We have sampled initial metallicities in the
range $0 \le Z/Z_{\odot} \le 0.1$, as representative of different enrichment 
histories, which may vary depending on metal diffusion and mixing. 

Neglecting differential enrichment by metals and dust grains, the initial
composition of the collapsing pre-stellar gas clouds directly reflects the
nucleosynthetic products of the polluting SN, its specific dust depletion
factor, relative gas-phase metal abundances and dust grain species and sizes.
We have therefore modified the model to allow a self-consistent 
description of metal-line cooling, dust thermal emission, dust-gas collision 
rate, H$_2$ formation rate on dust grains. In all theoretical studies
carried out so far, these processes were implemented under the implicit assumption
that the gas-phase metal abundances and dust grain properties were the same as 
those observed in the local interstellar medium. The only exception is the
study of Bromm \& Loeb (2003), where separate critical abundances of CII and OI
enabling cooling and fragmentation in collapsing gas clouds were derived. 
Still, they limited their study to the thermal evolution in the low-density regime,
including only metal line cooling and fragmentation.

The main results of our study can be summarized as follows:
\begin{enumerate}

\item As long as thermal physics is considered to play a major role in
setting a characteristic stellar mass scale, dust-induced fragmentation appears to be
the most promising mechanism for the formation of solar and sub-solar mass fragments. 
Since dust can be promptly synthesized in the ejecta of primordial SN, it is likely that its
contribution to thermal and fragmentation history of collapsing clouds at high redshifts 
had been relevant. 

\item Due to the strong depletion of C atoms onto AC grains and CO molecules for most SN models,
our analysis shows that CII makes a negligible contribution to fine-structure line cooling, leaving
OI as the main cooling agent in the gas phase.

\item Line-cooling, being limited to the initial evolution of collapsing gas clouds, for
densities in the range $10^4 - 10^6~$cm$^{-3}$, induces fragmentation into 
relatively large clumps, with characteristic fragment masses $M_{\rm frag} \geq 100 \Msun$,
even for initial gas metallicities $Z \lsim 10^{-2} Z_{\odot}$.

\item The formation of solar or sub-solar mass fragments
might operate already at metallicities $Z_{\rm cr} = 10^{-6} Z_{\odot}$.
The results that we find suggest a re-formulation of the critical metallicity scenario: 
indeed, once a self-consistent model is
adopted, the presence of gas-phase metals can not by itself lead to solar or sub-solar 
characteristic masses whereas
primordial supernovae of all masses (that is both in the standard SNII mass range 
and in massive PISN), appear to deplete enough metals in solid grains to enable the 
formation of low-mass fragments.

\end{enumerate}

Our approach to determine the minimum fragment masses as the Jeans masses corresponding
to the last fragmentation (temperature dip and $\gamma = 1$ condition) may be an
oversimplification of the complex dynamical evolution of cloud cores. 
Still, our findings are supported by results of 
a semi-analytical model for the linear evolution of non-spherical deformations by Omukai \etal
(2005) and by three-dimensional hydrodynamical simulations of Tsuribe \& Omukai (2006),
which confirm that fragmentation occurs during the dust-cooling 
phase and that sub-solar mass fragments are indeed produced. 

Let us emphasize that other processes might be equally important in reducing the
characteristic mass scales of metal-free or very metal-poor star forming regions.
Among these, the bimodal IMF suggested by Umemura \& Nakamura (2002), the presence
of a strong UV field in the vicinity of a very massive star (Omukai \& Yoshii 2003), 
HD cooling triggered in relic HII regions around very massive stars after the death 
of the exciting star (Uehara \& Inutsuka 2000; Nagakura \& Omukai 2005; Johnson \& Bromm 2006), 
shock-compression
induced by the first SN explosions (MacKey, Bromm \& Hernquist 2003; Salvaterra, Ferrara \& Schneider 2004).

Observationally, stellar relics in the halo of our Galaxy represent fundamental 
records of low-mass star formation at high redshifts. 
The HK-survey, Hamburg-ESO Prism Survey and Sloan Digital Sky Survey provide 
already a sample of about 5000 stars with [Fe/H]$<-2$, among which 750 with [Fe/H]$<-3$. The
project SEGUE (Sloan Extention for Galaxy Understanding and Exploration) shoud identify about
25000 stars with [Fe/H]$<-2$ (Beers \& Christlieb 2005). 
The interpretation of shape and low-metallicity tail of the metallicity distribution function shoud
provide important insights on the origin of the first low-mass stars that have formed in the Universe
(Schneider \etal 2003; Tumlinson 2006; Salvadori, Schneider \& Ferrara 2006). 

The surface elemental abundances of the observed stellar relics should shed light on the IMF of the
stars mostly responsible for early metal enrichment (Tumlinson, Venkatesan \& Shull 2004; Iwamoto \etal 2005;
Beers \& Christlieb 2005). In principle, dust depletion does not affect the interpretation of the observed chemical 
abundances because, once sufficiently high central densities are reached, dust grains present in collapsing
pre-stellar clouds are destroyed and the original gas-phase metal abundances (reflecting the nucleosynthetic
products of the SN progenitor star) are re-established. An alternative scenario has been recently proposed by
Venkatesan, Nath \& Shull (2006), who suggest that the elements forming the composition of the dominant dust 
compounds created in primordial supernovae (Si, Mg, O, C) can be selectively transported in the UV radiation field 
from the first stars and decoupled from the background SN metals in the gas phase. 
Interestingly, these elements appear to have enhanced abundances in the most metal poor halo stars observed.

\section*{Acknowledgments}
We thank S.Bianchi, H. Hirashita \& T.Kozasa for providing some of the data on grain optical
properties in a machine readable form and S.Bianchi for useful comments on the manuscript.

\label{lastpage}

\end{document}